# The HERMES-Technologic and Scientific Pathfinder

F. Fiore[a], L. Burderi[b], M. Lavagna[c], R. Bertacin[d], Y. Evangelista[a,w], R. Campana[a], F. Fuschino[a], P. Lunghi[c], A. Monge[e], B. Negri[d], S. Pirrotta[d], S. Puccetti[d], A. Sanna[b], F. Amarilli[f], F. Ambrosino[a], G. Amelino-Camelia[g], A. Anitra[h], N. Auricchio[a], M. Barbera[h], M. Bechini[c], P. Bellutti[i], G. Bertuccio[J], J. Cao[k], F. Ceraudo[a], T. Chen[k], M. Cinelli[l], M. Citossi[m], A. Clerici[n], A. Colagrossi[c], S. Curzel[c], G. Della Casa[m], E. Demenev[i], M. Del Santo[a], G. Dilillo[m], T. Di Salvo[h], P. Efremov[o], M. Feroci[a,w], C. Feruglio[a], F. Ferrandi[j], M. Fiorini[a], M. Fiorito[c], F. Frontera[p,a], D. Gacnik[q], G. Galgóczi[r], N. Gao[k], A.F. Gambino[h], M. Gandola[j], G. Ghirlanda[a], A. Gomboc[o], M. Grassi[s], C. Guidorzi[p], A. Guzman[t], M. Karlica[o], U. Kostic[n], C. Labanti[a], G. La Rosa[a], U. Lo Cicero[a], B. Lopez Fernandez[e], P. Malcovati[s], A. Maselli[d], A. Manca[b], F. Mele[j], D. Milánkovich[u], G. Morgante[a], L. Nava[a], P. Nogara[a], M. Ohno[r], D. Ottolina[c], A. Pasquale[c], A. Pal[v], M. Perri[a,d], R. Piazzolla[a], M. Piccinin[c], S. Pliego-Caballero[t], J. Prinetto[c], G. Pucacco[l], A. Rachevski[w], I. Rashevskaya[w,x], A. Riggio[b], J. Ripa[r,y], F. Russo[a], A. Papitto[a], S. Piranomonte[a], A. Santangelo[t], F. Scala[c], G. Sciarrone[a], D. Selcan[q], S. Silvestrini[c], G. Sottile[a], T. Rotovnik[q], C. Tenzer[t], I. Troisi[c], A. Vacchi[m], E. Virgilli[a], N. Werner[y,r], L. Wang[k], Y. Xu[k], G. Zampa[w], N. Zampa[w,m], G. Zanotti[c]

[a] INAF via del Parco Mellini 84, I00136 Roma, Italy; [b] Dipartimento di Fisica Università degli Studi di Cagliari, Italy; [c] Dipartimento di Scienza e Tecnologia Aerospaziali, Politecnico di Milano, Italy; [d] Agenzia Spaziale Italiana, Italy; [e] DEIMOS, Spain; [f] Fondazione Politecnico Milano, Italy; [g] Università Federico II, Napoli, Italy; [h] Dipartimento di Fisica e Chimica, Università degli Studi di Palermo, Italy; [i] Fondazione Bruno Kessler, Italy; [j] Dipartimento di Elettronica, Informazione e Bioingegneria, Politecnico di Milano, Italy; [k] Institute of High Energy Physics, Chinese Academy of Sciences; [l] Dipartimento di Matematica, Università di Roma Tor Vergata; [m] Dipartimento di Scienze Matematiche Informatiche e Fisiche, Università di Udine, Italy; [n] Aalta Lab, Slovenia; [o] University of Nova Gorica, Slovenia; [p] Dipartimento di Fisica e scienze della Terra, Università di Ferrara; [q] Skylabs, Slovenia; [r] ELTE - Eötvös Loránd University, Hungary; [s] Università di Pavia, Italy; [t] IAAT, EKUT - Eberhard Karls Universität Tübingen, Germany; [u] C3S, Hungary; [v] Konkoly Observatory, Hungary; [w] INFN; [x] TIFPA-INFN; [y] Department of Theoretical Physics and Astrophysics, Masaryk University, Brno, Czech Republic

## ABSTRACT

HERMES-TP/SP (High Energy Rapid Modular Ensemble of Satellites Technologic and Scientific Pathfinder) is a constellation of six 3U nano-satellites hosting simple but innovative X-ray detectors, characterized by a large energy band and excellent temporal resolution, and thus optimized for the monitoring of Cosmic High Energy transients such as Gamma Ray Bursts and the electromagnetic counterparts of Gravitational Wave Events, and for the determination of their positions. The projects are funded by the Italian Ministry of University and Research and by the Italian Space Agency, and by the European Union's Horizon 2020 Research and Innovation Program under Grant Agreement No. 821896. HERMES-TP/SP is an in-orbit demonstration, that should be tested starting from 2022. It is intrinsically a modular experiment that can be naturally expanded to provide a global, sensitive all sky monitor for high-energy transients.

# 1. INTRODUCTION

The coalescence of compact objects, neutron stars (NS) and black holes (BH), and the sudden collapse to form a BH, holds the keys to investigate both the physics of matter under extreme conditions, and the ultimate structure of space-time. At least three main discoveries in the past 20 years prompted such studies.

First, the arcmin localization of Gamma-Ray Bursts (GRBs, sudden and unpredictable bursts of hard-X/soft γ-rays with huge flux up to $10^{-3}$ ergs/cm$^2$/s), enabled for the first time by the instruments on board BeppoSAX, allowed to discover their X-ray and optical afterglows [1,2], which led to the identification of their host galaxies [3]. This definitely confirmed the extragalactic nature of GRBs and assessed their energy budget, thus establishing that they are the most powerful particle accelerators in the Universe. Even accounting for strong beaming, the energy released can indeed attain $10^{52-53}$ erg, a large fraction of the Sun rest mass energy, in ≈0.1–100 seconds, produced by bulk acceleration of plasmoids to Lorentz factor Γ≈100–1000 (see e.g. [4,5]).

Second, the large area telescope (LAT) on board the Fermi satellite established GRBs as GeV sources [6], confirming their capability to accelerate matter up to Γ≈100–1000 and allowing us to apply for the first time the program envisioned by Amelino-Camelia and collaborators at the end of the '90s to investigate quantum space-time using cosmic sources [6,7].

Third, August 2017, with the GW170817 event, marked the beginning of the so-called *multi-messenger* astrophysics, in which new observations of Gravitational Waves (GW) added up to traditional electromagnetic observations from the very same astrophysical source [8]. The nearly simultaneous detection of a GW signal from merging NSs – or merging of a NS with a BH – and of a short Gamma-Ray Burst detected by the Advanced LIGO/Virgo the former and the Fermi and INTEGRAL satellites the latter [8], proved that short GRBs can be associated with merging NSs, as hypothesised since about thirty years. Furthermore, the quick localisation of the GW-GRB event produced the discovery and the detailed study of a kilonova event, powered by the radioactive decay of r-process nuclei synthesized in the ejecta [9,10]. During the past >20 years from the localisation of the first GRBs[1,2,3], no kilonova was unambiguously discovered or spectroscopically studied, and these elusive *but* crucially important events for the synthesis of rare heavy elements remained a beautiful but unproven theoretical idea[11].

In the next few years Advanced LIGO/VIRGO will reach their nominal sensitivity and more gravitational interferometers will come online (KAGRA in Japan and LIGO-India). **The operation of an efficient X-ray all-sky monitor with good localisation capabilities will have a pivotal role in bringing multi-messenger astrophysics to maturity**, and to fully exploit the huge advantages provided by adding a further dimension to our capability to investigate cosmic sources. The HERMES (High Energy Rapid Modular Ensemble of Satellites) program offers a fast-track and affordable complement to more complex and ambitious missions for relatively bright events. This paper is organized as follows. In Sect. 2 we discuss the science behind next-generation sensitive all-sky-monitor. In Sect. 3 we present our implementation strategy and provide a broad description of the HERMES Technologic and Scientific Pathfinder projects. Finally, in Sect. 4 we put our projects in the context of the on-going initiatives on transients monitoring.

# 2. THE SCIENCE OF HERMES

A further increase of the discovery space on the physics/astrophysics of high-energy transients, and on the use of transients as tools to search for new physics, can lead to breakthrough discoveries in at least three broad areas:
1) High-energy Transient localization
2) The GRB inner engine
3) The granular structure of space-time

We will discuss the three topics in the following sections.

## 2.1 High-energy Transients localization

The combined Advanced LIGO/VIRGO error-box of the GW170817 event was of the order of 30deg$^2$[8,9,10], within the >10 times larger Fermi/INTEGRAL error-box of GRB170817. The Advanced LIGO/VIRGO detection indicated a very close event, ~40 Mpc, greatly limiting the volume, and therefore the number of possible targets (about 50 galaxies were present in the LIGO/VIRGO error-box within 40Mpc). For fainter events, likely further away, such as those that will be provided by Advanced LIGO/VIRGO after the refurbishment following Observation Runs O3, and by the next generation interferometers during the 2020ties, the volume to be searched will be much larger: the horizon for NS-NS merging events detected with a signal-to-noise of 8 will reach ~200 Mpc for LIGO and 100–130 Mpc for VIRGO in a few years, implying a discovery volume ~100 times larger than in the GW170817 case. In such volumes the number of optical and near infrared transients may be as large as hundreds or even thousands, making difficult the prompt identification of the GWE optical/IR counterpart for further follow-up. Fortunately, the number of X-ray transients in the same volume is much more limited (from zero to a few), making much more efficient the search for the electromagnetic counterparts of GWEs in this energy band.

In addition, the study of sources in the 20 GeV–300 TeV energy range will experience a major boost starting from the first half of the 2020ies thanks to the Cherenkov Telescope Array (CTA), an array of new generation Cherenkov telescopes located in two different sites, covering both the southern and northern hemisphere. Construction of the full array is expected to end in 2025. Three GRBs have been detected at TeV energies so far, one by MAGIC [12] and two by HESS [13,13b]. The CTA fast re-positioning capabilities (20 seconds) and the largely improved sensitivity and low energy threshold (~20GeV) as compared to MAGIC and HESS makes the study of GRB high-energy radiation very appealing, opening a window on the possibility to accurately derive the jet Lorentz factor, the role of synchrotron and inverse Compton radiation, and the magnetic field strength and configuration. Given the limited field of view (FoV) of CTA at GeV energies (4.3°), an instrument operating from the beginning of the '20s and providing localization of GRBs with error smaller that the CTA FoV is of paramount importance to trigger the CTA follow-up observations.

Today, four active satellites provide GRB and X-ray transients positions with accuracy from several arcmin (Swift/BAT, INTEGRAL/IBIS, AGILE/Super-AGILE), to many degrees (Fermi/GBM). In addition, for events with localization uncertainties > several degrees, detectors on board different satellites can be used jointly in the Inter-Planetary Network (IPN) to provide error circles in the sky with typical width a few degrees, reaching a few arcmin is rare cases. All these satellites were launched in the 2000s, being operational for more than a decade. It is not guaranteed they will keep being operational up to the 2020s on. For this reason, several proposals to NASA and ESA have been already submitted to select the successor of these instruments (e.g. ESA-M5, NASA-Explorer, NASA Decadal Survey). Even if approved, these missions are based upon complex large-area detectors in a single spacecraft, with a total cost of a few hundred million Euro and a development timescale of at least 10 years, and therefore will hardly be ready before the end of the 2020s. *HERMES may complement such more ambitious missions for bright transients, bridging the gap between current X-ray monitors and the next generation* (see Section 4).

The way in which HERMES determines the transient position has been used since the VELA pioneering observations during the '60s and '70s. The position of the transient is obtained by studying the delay between the arrival time of the signal upon different detectors, placed hundreds/thousands of km away. The accuracy in determining the position ($\sigma_{PA}$, position accuracy) can be approximated by the following simple formula:

$$\sigma_{PA} \approx \frac{\sqrt{\sigma_{delay}^2 + \sigma_{TPOS}^2 + \sigma_{TIME}^2 + \sigma_{sys}^2}}{AB\sqrt{N-1-2}} \qquad eq.1$$

where $\sigma_{delay}$ is the error on the delay obtained applying cross-correlation techniques between two light curves, $\sigma_{TIME}$ is the uncertainty on absolute time of the collected photons, $\sigma_{TPOS} = \sigma_{POS}/c$ is the error due to the uncertainty on the position of the satellite, $\sigma_{sys}$ is a systematic uncertainty, AB is the Average Baseline, namely the average distance between detectors, and N is the number of detectors observing a given event simultaneously. $\sigma_{TPOS}^2 + \sigma_{TIME}^2 < \sigma_{delay}^2 + \sigma_{sys}^2$, as expected in Low Earth Orbit (LEO), where accurate timing (a few hundreds of nanoseconds) and accurate positions (tens of meters) can be achieved thanks to the GPS system. Eq. 1 can then be simplified to:

$$\sigma_{PA} \approx \frac{2.4° \sqrt{\sigma_{delay}^2 + \sigma_{sys}^2}}{\sqrt{N-1-2}} \qquad eq.2$$

where AB has been assumed 7000 km and $\sigma_{delay}$ and $\sigma_{sys}$ are given in ms. *This means that if $\sigma_{delay}$ and $\sigma_{sys}$ can be kept of the order of 1ms or less, the accuracy reachable in principle even by a small number of detectors is the order of a few degrees or less*. $\sigma_{delay}$ depends on: a) the temporal resolution of the detector; b) the temporal structure of the transient; c) number of photons per temporal bin in the light-curve, which is basically given by the strength of the transient and by the collecting area of the detector; and d) the level of the background, see [14] for a more accurate discussion on the determination of the transient position. The residual systematic error is more difficult to estimate. Contributions may come from: a) fraction of the Earth atmosphere in the FoV of the detectors, since the transient signal can be reflected and smoothed by the atmosphere; b) detection of the transient during large background variations; c) detection of the transient at different off-axis angles in different detectors coupled with an insufficient calibration, etc. One of the goals of the HERMES pathfinder experiment is to better study all these, and potential other systematic uncertainties.

## 2.2 The GRB inner Engine

GRBs are thought to be produced by the collapse of massive stars and/or by the coalescence of two compact objects. Their main observational characteristics are the huge luminosity and fast variability, often as short as 1ms or even less [15,16]. These characteristics led to the development of the fireball model, i.e. a relativistic bulk flow where shocks efficiently accelerate particles. The cooling of the ultra-relativistic particles then produces the observed X-ray and gamma-ray emission. Further interactions of the relativistic wave fronts with the Interstellar Medium then produce the so-called afterglow emission, seen from radio to γ-ray energies. While successful in explaining GRB observations, the fireball model implies a thick photosphere, hampering direct observations of the hidden inner engine that accelerate the bulk flow. We are then left in the frustrating situations where we see at work daily the most powerful accelerators in the Universe, but we are kept in the dark over their operation. One possibility to shed light on their inner engines is through GRB fast variability. Early numerical simulations [17], as well as modern hydro-dynamical simulations [18], and analytic studies [19] suggest that the GRB light-curves reproduce the activity of their inner engines. GRB light-curves have been investigated in detail down to 1 ms or slightly lower [15,16], while the μs–ms window is basically unknown, as little known is also the real duration of the prompt event. We do not know how many shells are ejected from the central engine, which is the frequency of ejection as well as its length. *Accurate and sensitive GRB timing including the μs−ms window should help in answering at least some of these questions.*

One of the most important open questions in GRBs is the composition of the jets. It is still not clear, indeed, whether the jet energy is mostly carried out from the central engine in the form of Poynting flux (magnetic jet) or as kinetic energy of the matter (baryonic jet). In the two scenarios, the mechanism for extraction of the jet energy is very different, being dominated by magnetic reconnection events in the first case, and by internal shocks in the second case. The two scenarios predict different variability properties. Fast variability should be related to the location where dissipation takes place in the matter dominated model, while in the magnetic model it reflects the possible presence of turbulent regions with fast proper motions within the same emission region. *The detection of sub-millisecond variability is then fundamental to place constraints on both models and on the composition of the jet itself.*

The nature of the radiative mechanism responsible for the observed GRB prompt radiation is still uncertain. Recent analyses combining observations from hard X-ray (Swift/BAT and Fermi/GBM) and soft X-ray (Swift/XRT) instruments have revealed the presence of a new feature in many spectra: a spectral break at several keV, below which the spectrum hardens. The overall broadband behavior is in these cases consistent with synchrotron radiation and it provides strong constraints on the location and properties of the emission regions. These findings proved the importance of extending systematically the observations a few keV to a few MeV. Today this is feasible only for a small number of GRBs, because the XRT starts observing after 60–100 seconds after the GRB trigger, when its prompt emission has usually ended. *To overcome these difficulties, the next generation of sensitive all sky monitors should provide simultaneous observations in the keV–MeV broad-band.*

Fast timing, at the level of a fraction of μs, is at the heart of the HERMES concept. While HERMES is not a monolithic detector, once the position of the GRB is inferred by time delays of the signal detected at different detectors, the same signals can be realigned in time and added together to improve the statistics of the single detectors. The total collecting area needed to provide a few counts per 10μs for GRBs with a photon count rate of $10 ph/s/cm^2$ in the energy band 50–300 keV is a few $m^2$, that translates into ~50–100 detectors looking simultaneously at the source, each having a collecting area of ~$200 cm^2$. The assumed photon count rate corresponds to a fluence of $4\times10^{-5}$ erg/$cm^2$, assuming a burst of duration 20s and a 50-300 keV power law spectrum with $\alpha_E$=0.6. The total number of GRB brighter than this fluence is about 2/month on the full sky. It should be noted that both a positive correlation between luminosity and variability, and a negative correlation between the width and the amplitude of each single pulse in a burst have been suggested by [20] using BATSE data, both trends favoring the search for short time variations in bright bursts. In addition to astrophysics studies, fast GRB variability and large collecting areas can also be used for fundamental physic studies, as discussed in the next section.

## 2.3 The granular structure of space time

Several theories proposed to describe quantum space-time, predict a discrete structure for space-time on small scales, $l_{min} \sim l_P = [G\hbar/c^3]^{0.5} \sim 10^{-33}$cm. For a large class of these theories this space discretization implies the onset of a dispersion relation for photons, that could be related to the possible violation or deformation of the Lorentz invariance on such scales. Special Relativity postulate Lorentz invariance: all observers measure the same speed of light in vacuum, independent of photon energy, which is consistent with the idea that space is a three-dimensional continuum. On the other hand, if space is discrete on very small scales, it is conceivable that light propagating in this lattice exhibits a sort of dispersion relation, in which the speed of photons depends on their energy [7,21,22 and references therein].

When two photons of different energies $E_2 - E_1 = \Delta E$ emitted at the same time travel over a distance $D(z)$, because of this dispersion relation, they exhibit a delay $\Delta t$ in their arrival times. The predicted quantum-space-time effects modifying the propagation of light are extremely tiny, but they could cumulate along the way, e.g for signal traveling over cosmological distances, and might be revealed by observing hard/soft spectral delays. GRBs, being so powerful and distributed uniformly at cosmological distances, are the best phenomena to test quantum gravity [7]. Such a detection could directly reveal, for the first time, the deepest structure of quantum space-time by gauging its structure in terms of the photon energies. The variation of arrival time with energy is roughly given by $\Delta t \approx \frac{\Delta E}{E_{QG}} \times D(z)/c$ where $E_{QG}$ is the energy at which quantum space-time effects become relevant. Fermi LAT detection of the short GRB090510 at GeV energies where used to put constraints on $\Delta t$ [23]. They find $\frac{\Delta t}{\Delta E} \leq 30$ms/GeV, putting a constraint on $E_{QG}$ of the order of the Planck energy. This limit, however, is obtained by assuming that a single observed 31 GeV photon was emitted simultaneously to other GeV photons, making a ~0.2s burst. A much more robust limit would had been obtained by measuring the delay between two well defined burst shapes at low and high energies, i.e. by constraining the cross-correlation function between two low and high energy light-curves. Unfortunately, this is well beyond the capability of Fermi LAT. One possibility to increase the number of photons, thus sampling better GRB light curves, is to make observations at lower energies. However, this requires constraining much shorter timescales. As an example, to reach an accuracy on $E_{QG}$ of the order of $E_{Planck}$ using 100 keV photons instead of GeV photons would require pushing the timing to 3μs against 0.2s. [23] have applied the cross-correlation method to a sample of 21 Swift BAT short GRBs finding a limit on $E_{QG}$ of the order of 0.001 $E_{Planck}$. To improve this limit by a factor ~1000 we need a statistical sample of bright long GRBs with ms or, even better, sub-ms temporal structure, large collecting areas (at least a factor of 10 larger than Swift BAT, that is a few/several m² at 100keV), higher energy response, up to at least 1 MeV. *A reasonable limit for a new large hard-X-rays, soft gamma-ray facility is thus:*

$$\frac{\Delta t}{\Delta E} \leq \frac{3\mu s}{100 keV} = \frac{30 \mu s}{1 MeV}$$

*which will allow probing the Planck scale* (see [22] for more details).

## 3. IMPLEMENTATION STRATEGY

The HERMES approach is different from that commonly adopted for space-based observatories. HERMES is an experiment distributed over several units. The main advantages of the HERMES strategy with respect to standard High-Energy Astrophysics experiment are two: modularity and quick development with limited costs.

Modularity allows:

a. to firstly fly a reduced versions of HERMES (the HERMES pathfinders) to prove the concept in space;
b. to avoid single (or even multiple) point failures: if one or several units are lost the constellation and the experiment can still be operative;
c. to fully test the hardware in orbit with the first launches and then improve it, if needed, with the following launches (*iterative development*).

The HERMES project development incremental strategy is showed in figure 1. HERMES will exploit commercial off-the-shelf (COTS) hardware, as well as the trend in reducing the cost of both manufacturing and launching nano-satellites over the next years. HERMES would naturally fit a scheme where mass-production of identical units would follow the development and testing of single test units. Iterative and incremental development processes followed by mass-production are certainly not new in the New Space panorama, HERMES constitutes however the first systematic application of these concepts to a mission with the ambition of achieving breakthrough science in the very competitive fields of astrophysics and fundamental science.

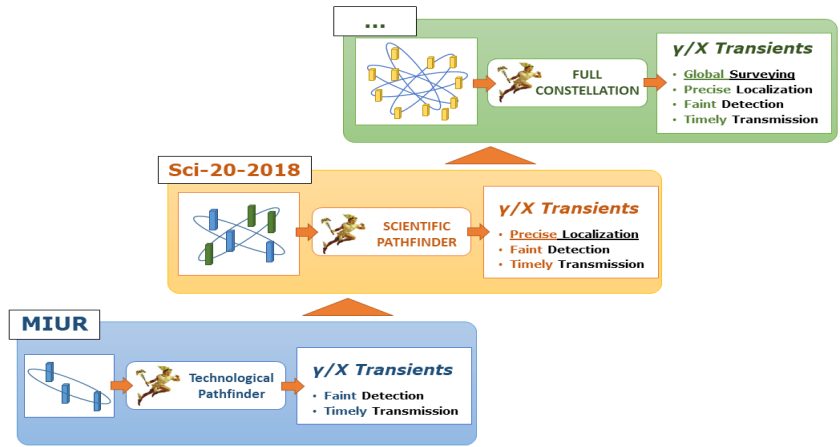

Figure 1. HERMES project philosophy incremental development

**3.1 Objectives of the HERMES Full Constellation**

Based on these scientific goals presented in Section 2, the HERMES Full Constellation (HFC) should be able to reach the following high-level scientific requirements:

1. *All sky coverage*. The full sky should be accessible at each given moment. Every point should be seen by at least 20-30 detector at an off-axis angle $\leq 60°$.

2. *50–300 keV flux sensitivity of ~0.2 ph/cm$^2$/s for short GRBs*. This would allow detection of events similar to GRB170817 up to ~120–150Mpc, the expected sensitivity horizon for binary NSs of Advanced Virgo during O4 and at the beginning of O5. As a comparison, 0.1 ph/s/cm$^2$ is the current sensitivity of Swift/BAT in the 15–150 keV band, but on a limited 1.4sr FoV, while the sensitivity of Fermi/GBM in the 50–300 keV band is ~1ph/cm$^2$/s.

3. *Accuracy of localization of bright long GRBs better than 15′* on each coordinate, corresponding to a box of error of area <0.2deg$^2$. This would allow prompt identification of the transient counterpart at different wavelengths, greatly easing the multi-wavelength follow-up.

4. *Accuracy of localization of short GRB better than 1°* in both coordinates, corresponding to a box of error of ~3deg$^2$. This would allow prompt identification of the electromagnetic counterpart of a GWE and guide prompt multi-wavelength follow-up facilitating the identification of the event's host-galaxy. In fact, the number of galaxies to be searched for a transient in the case of the GW170817 event was about 50 in an error box of ~30deg$^2$. To keep about constant this number while increasing by a factor of ~10 the volume, requires a shrinking by the same factor of the error box size.

5. *Determination and dissemination of the transient position within minutes*. GRB afterglows decay rapidly, hence, the prompt dissemination of accurate GRB positions is key for an optimal multi-wavelength follow-up.

6. *Accuracy of timing better than a few hundred nanoseconds*, to provide errors on the delays obtained via CCF between GRB signals on different detectors < of a few tens µs and to investigate the sub-ms window.

7. *Total collecting area of the order of a few m$^2$*. This would allow feeding light curves of relatively bright GRBs (>10ph/s/cm$^2$) with a few counts every 10 µs, thus allowing to fully exploit the good time accuracy.

**3.2 Objectives of the Pathfinder programs**

The pathfinder programs must prove in space the concept and the technical feasibility of the HERMES approach. The Italian Space Agency (ASI) approved on 2016 a first pilot project to study innovative, compact X-ray detectors optimized for GRB observations. The second project approved was the HERMES Technologic Pathfinder (HERMES-TP) project, funded by the Italian Ministry for education, University and Research (MIUR) and ASI. HERMES-TP is aimed at producing three full units (payload + service module). The third project approved was the HERMES Scientific Pathfinder

(HERMES-SP), funded by the European Union's Horizon 2020 Research and Innovation Program under Grant Agreement No. 821896. HERMES-SP is aimed at producing three additional complete units and at developing the mission and science operation centers (MOC & SOC). Finally, ASI approved and funded the participation to the project SpIRIT (Space Industry Responsive Intelligent Thermal), founded by the Australian Space Agency (ASA), and led by University of Melbourne. SpIRIT will host an HERMES-like detector and S-band transmission systems. The HERMES-TP/SP mini-constellation of six satellites plus SpIRIT should be tested in orbit during 2022. Launch of the HERMES-TP/SP six units will be provided by ASI.

The main goals of HERMES-TP/SP are:

1. *Develop miniaturized scientific instrumentation and technologies for breakthrough science*: HERMES-TP/SP is developing miniaturized X-ray detectors to catch signals from GRBs and other X-ray transients with the challenge to fit them in a nano-satellite, still preserving the level of sensitivity and performances required for breakthrough science.

2. *Prepare for relevant scientific data production:* HERMES-TP/SP will demonstrate that accurate GRB positions can be obtained by measuring the delay time of arrival of the signal to different detectors. The in-orbit demonstration is aimed at achieving position accuracies of the order of a few degrees, already better or comparable with that of the Fermi GBM. Arcmin position accuracies could then be achieved by HFC, when the same event is observed by tens/hundreds of detectors. The same excellent timing accuracy needed for temporal correlation analyses will also allow to investigate the temporal structure of high energy transients down to a fraction of μs.

3. *Demonstrate the COTS applicability to challenging space missions*: HERMES-TP/SP shall refer to miniaturised platforms to contain the costs and the time-to-orbit: CubeSats represent a cluster of disruptive technologies, risky at the time being because of COTS inadequacy to space environment. HERMES-TP/SP will assess and apply a production life cycle aimed at increasing the COTS reliability still limiting the time-to-orbit and the cost.

4. *Contribute to the Space 4.0 goals and expectations*: HERMES-TP/SP will contribute identifying and standardizing new and innovative approaches to manufacture, assemble and test miniaturized components, allowing to change the market mechanism in space sector, allowing more brilliant and proactive small and medium-sized enterprises (SMEs) to access the business.

5. *Enlarge and strengthen the space distributed architectures and mega-constellations applicability and reliability*: to gain the required precision in localisation and timeliness for an unpredictable physical event in space and place requires a sensors distribution covering the full sky. HERMES-TP/SP will investigate and optimize the space-ground segments nets design in terms of number of satellites vs number of ground stations, number of satellites vs the need of spare satellites, heterogeneous vs homogeneous orbits geometry, constellations architecture and coordination and space segments functionalities (e.g. science, relay, etc.), on board sub-systems performance, lifetime vs to replacement, space segment reliability, costs, time to launch, time to operation, scalability. The study will contribute deepening the knowledge on potentials and limitations for large scale -small platform distributed architectures applicability to challenging space science missions.

The high-level scientific requirements of the Pathfinder are the following:

- **Detect GRBs with peak flux ≥0.5–1 ph/s/cm$^2$ in the 50–300 keV band;**

- **Detect ≥40 long GRBs and ≥8 short GRBs simultaneously in at least 3 units with an efficiency ≥40–50% in each unit, during the nominal phase of 2 years,** to be able to assess their position through the analysis of the delay time in the signal arrival time on different detectors.

The first requirement implies a 50–300 keV collecting area of the single detector ≥50cm$^2$ and background ≤1.5 counts/cm$^2$/s [24,25,26]. The second requirement implies a large FoV for each detector (a few steradians), and the possibility to co-align the FoVs seen at every given time by at least three detectors. The accuracy of the attitude determination should be better than a small fraction of the detector FoV, say several degrees. A detailed mission analysis has been performed to verify this requirement can be met, see e.g. [27].

The scientific key goal of the pathfinders is to prove that accurate localizations can be obtained by miniaturized instrumentation on board nano-satellites, using the delays of the arrival time of the signal to detectors positioned at thousands km of distance. According to eq. 1 and eq. 2, the localization accuracy scales as the inverse of the average projected baseline, the inverse of the square root of the number of satellites observing simultaneously the GRB event and thus as the inverse of the square root of the total collecting area. Scaling the localization accuracy by a factor of ~5 for the observation multiplicity (3–4 pathfinder detectors seeing the same event instead of 25–30 for HFC) and a factor ~2 for the collecting area, means that to be able to reach the HFC requirement of ~15arcmin the pathfinder mini-constellation must demonstrate an accuracy on the localization of the order of 2.5°. This requires the ability to constrain the delay between two GRB light-curves with a ~1ms uncertainty or less, which in turn implies a collecting area of at least 50 cm$^2$ for the pathfinder detectors, to collect at least 1 counts every a few ms for relatively bright GRBs (>10ph/s/cm$^2$), see [14, 24]. The pathfinder mini-constellation is expected to fly on a single orbit (a LEO with height between 500 and 600 km and inclination ≤ 20°), therefore a precise transient localization will be limited to the Right Ascension coordinate only. Improvements on the Declination coordinate will be obtained by increasing the number of elements in the fleet and/or adding elements in orbital planes inclined with respect to the equatorial plane. SpIRIT will fly on Sun-synchronous orbit (SSO), thus providing a first element to constrain accurately also the transient declination.

The requirements on the payload are ambitious: broad energy band, good efficiency, good energy resolution, high temporal resolution, extremely compact design, reliable operation in a quite broad range of space environments (temperature, magnetic field, radiation, etc). These requirements are certainly reachable on standard scientific experiments, but to meet them on a CubeSat, with very limited weight and power resources is a challenge that can be met only through innovative technologies and approaches.

### 3.3 Payload

The HERMES-TP/SP payload is described in detail in [24, 25, 26, 28, 29, 30]. Here we recall the main concept, heritage and synergies. Given the required broad energy range, an integrated detector is adopted, exploiting a "double detection" mechanism, with a partial overlap of the two systems around ~20 keV, see Figure 2. Detection of soft X-rays is obtained by employing the Silicon Drift Detector (SDD) technique [31], with individual detectors having a size <1cm$^2$, which allow us to attain a low noise level (of the order of a few tens of e- rms at room temperature) and correspondingly a low threshold for the detection of X-ray radiation down to a few keVs. A 450 μm thick silicon detector is able to detect X-ray radiation with a good efficiency up to 20–30 keV. Detection of hard X-rays/gamma-rays is obtained with a two-stage process that first converts the photon energy into visible light produced by a scintillating material, which is then collected and converted into electric charge in a photodetector. The SDD is also used to directly detect X-ray photons; in this case, the SDD acts as a photodiode and produces an amplitude charge signal related to the amount of collected scintillation photons. The discrimination between the two signals in the SDD (soft X-rays or optical photons) is achieved using a segmented design: each scintillating crystal is seen by two SDDs, so that events detected uniquely by a SDD are most likely associated to soft X-rays hitting a single SDD, while events detected simultaneously by two SDDs are most likely associated to optical light produced in the scintillator by an incoming gamma ray. The crystal selected for the HERMES application is GAGG:Ce (Cerium-doped Gadolinium Aluminum Gallium Garnet). This rather new material is characterized by a fast response (well below 1 μs) and high light yield (~56 photons/keV), which allows reaching a lower energy threshold with respect to a more standard scintillator of similar density, such as Bismuth Germanate (BGO, ~8 photons/keV).

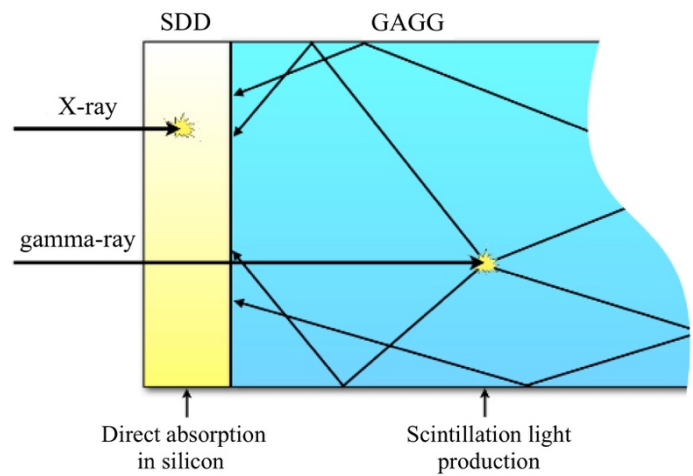

Figure 2. The siswich detector concept. SDDs are used to read out both the optical light from the scintillator crystal produced by a gamma-ray interaction, and directly the X-rays coming from cosmic sources.

GAGG crystals have not been routinely used in space application yet. Publications describing performances of GAGG in radiation environment confirm that GAGG crystal can be safely used in space [32] Furthermore, we have performed tests at facilities in Italy (Trento Protontherapy Centre, January 2019). Results confirm that GAGG can be safely used in space, as detailed in [33]. The exclusion of traditional photomultiplier readout is due to their limited efficiency (~20% vs. 90% of Si detectors), larger volume and much higher power consumption. A great advantage of SDDs compared to SiPM (Silicon Photomultipliers) lies in their ability to also directly detect low energy X-rays, thus allowing for a compact broadband instrument. Furthermore, SDDs have already shown a greater resistance to damage due to radiation compared to SiPM.

To take full advantage of the low noise characteristics of the SDD, the readout electronics must be optimized for the very low capacity of the charge collection anode. Furthermore, the electronics must be small and with a very low power consumption. These requirements led the choice of an ASIC (Application Specific Integrated Circuits) as the front-end electronic. Our baseline design includes two ASICs. The first one, the LYRA-FE, has the function of charge-sensitive pre-amplifier, directly connected with the SDD on the same board, to fully exploit their low noise; the second one, the LYRA-BE, includes all the functions of a spectroscopic chain, including the analog-to-digital converters (ADCs) and a fast discriminator for the timing of the events, see [29 for details]. The heritage for the analog front end electronic stems from the VEGA ASIC project, a mixed-signal low-noise multichannel ASIC developed by Politecnico di Milano and the University of Pavia for the readout of SDDs. The back-end electronics (BEE) board houses the second analog stage, the ADC and all the necessary control logic for the management of the signals. This board is also responsible for the management of the on-board-time, a key element of the HERMES approach. The baseline solution implemented in the BEE FPGA uses 2 counters: one incremented by the GPS pulse-per-second (PPS) and the other incremented by the 10 MHz clock of a miniaturized atomic clock. When the GPS signal is locked, the PPS signal is synchronized every second to the rising edge of the PPS provided by the satellite bus, thus ensuring a continuous correction for the payload on-board time. In case of GPS not locked, for example when the number of the GPS satellites tracked by the satellite bus GPS receiver is lower than 4 due to the earth occultation, the miniaturized atomic clock still provides the 1PPS and 10 MHz clock signal in a free-running mode. We studied an alternative solution to the atomic clocks, represented by temperature compensated crystal oscillators (TCXOs) and voltage-controlled temperature compensated crystal oscillators (VCTCXOs), which have the advantage of lower power consumption, smaller part height, and the redundancy in the form of having more than one of such parts to the control electronics at the expenses of some loss of performances. The BEE board receives the signals from the LYRA-BE and directly interfaces with the photon data handling unit (PDHU). The PDHU is described in detail in [30] is composed by the following functional parts: 1) Interface with the Power Supply Unit (PSU); 2) Interface with BEE; 3) Interface with satellite bus; 4) Interface with GPS; 5) Payload CPU; 6) Payload memory. The PDHU operations will consist of 1) Operative mode management; 2) Photon list generation & FIFO events buffering; 3) Photon list 'cleaning' for noisy events (particle); 4) Burst trigger logic management; 5) Scientific Data packet formatting; 6) Housekeeping (HK) management.

The HERMES pathfinder payload mass, power and data budget are presented in detail in [24]. The mass is <1.6 kg, power consumption is <6W and the amount of data generated daily is <1 Gbit/day.

### 3.4 System development philosophy

The critical aspect of the CubeSat approach lies primarily in the high mortality of nano-satellites as soon as they are inserted in orbit. The main reasons why they stop working are the use of miniaturized components, typically commercial-off-the-shelf (COTS), not space-qualified, and a design prone to single point failure and often poor in redundancy. *The HERMES pathfinder project wants to revisit the CubeSat production risky philosophy*, which leads to a very low, unacceptable level of product return, toward a more reliable and robust production process, identifying the correct balance of the build-test-verify–launch approach against the currently preferred build-launch strategy to produce reliable platforms which can serve real scientific payloads for long lasting space missions with relevant data return.

To minimize the risk of mortality for the HERMES Pathfinder mission, an incremental strategy for the development of the flight units is adopted, paying particular attention to the design, characterization, testing and verification phases of the first unit, the Proto Flight Model (PFM); the PFM will be used as Flight Model # 1 (FM1). In order to obtain a robust solution, from the design phase, engineering models or breadboards or beta versions of HW and SW critical subsystems/components have been used. These models/breadboards were realized and tested early in the project, and were used to improve and finalize the design and the integration procedures in an iterative approach. We then concentrated the effort in the development of the PFM. This approach will be critical for the success of the mission. The use in the design phase of a rapid prototyping and testing of parts of the system extends the time dedicated to the definition of the system, but greatly increases reliability, and, on the other hand, the possibility of managing any anomaly that may arise in flight, thanks to the detailed knowledge of the behaviour of the on-board components in the operating environment. The PFM will therefore be subject to functional tests at both subsystem and system level. Communication tests using both the VHF/UHF and S-band system with a ground station will also be performed to demonstrate the proper functioning of the radio link for a given distance between the ground station and the satellite. Finally, the PFM will be subject to environmental tests (thermal vacuum and random vibration) at proto-flight levels.

Such approach is currently not applied to the nano-satellites production cycle: the characterisation, verification and testing processes of both software and hardware for single components and assembly typically accomplished for regular satellites are withdrawn in CubeSat realisation to speed up the time to flight and minimize the costs.

The remaining units will be produced at FM level, and tested at system level only, to stay in the overall cost cup and speed up the time to launch. Parallelization of activities will be carried out for the production and test of the payload and SM FM. This activity will also be preparatory to imagine and design the assembly line that will be necessary to put into practice if we want to realize a full constellation of tens/hundreds of units, which cannot clearly be implemented and tested following a traditional logic.

### 3.5 Service module

The Service Module (SM) selected for the HERMES-TP/SP project is a CubeSat of the 3U class. It offers a volume of 10×10×30 cm and a total mass of the order of 5–6 kg. The small size associated with a low cost for the construction of each vehicle, make this solution optimal in the context of the proposed scientific project, allowing the deployment in orbit of large number of replicating sensors, in short time and at sustainable costs, and to answer the localisation functionality goal of the project.

The SM can be broken down on well-known on-board s/s: Structure & mechanisms; Electric Power s/s (EPS); thermal Control s/s (TCS); Telemetry, Tracking & Command s/s (TTC); Attitude and Orbit Determination and Control s/s (AODCS); On Board Data Handling s/s (OBDH). The philosophy is based in high technology readiness level (TRL) components selection: possibly in space, if already flown, otherwise high in their applicative domain: in this case a dedicated verification & validation plan will be identified through a similarity analysis process to map the current Earth TRL into the Space technology TRL scale and trade off the assembly, integration, test, verification (AIT/AIV) related approach. The main design driver will be to ensure, at the minimum, the space segment basic functionalities whenever non-nominal conditions occur, to be robust to mortality, typically occurring because of lack of electronics: the survive and communicate functionalities will be mandatory even before science operations accomplishment. Therefore, the EPS and TTC s/s drove the design.

Because of a continuously operating detector, even in standby mode, and the need to be continuously ground-connected for scientific events occurrence communication, the electric power resource may represent a criticality: CubeSat in fact stay in the order of few watts supply, because of the limited surface to body mount solar cells. To increase the available power to 20–30watts two deployable wings are included in the SM. The TCS will be passive at the most, to save power request on board; TTC will focus on the science data robust transmission, taking advantage of the IRIDIUM module to ensure RF connection anytime a GRB event may occur; download of scientific data will be done using an S-band transceiver/antenna system. To ensure robustness, omnidirectional antennas will be mounted on board supporting UHF/VHF connections, to protect the mission loss from attitude control potential anomalies. No orbital control is foreseen for the application, while the state vector determination will benefit of the GPS receiver and the inertial measurement unit (IMU) for the centre of mass, and magnetometers and sun sensors for the attitude; attitude will be controlled, by means of magneto-torquers and reaction wheels for the platform health and operability maintenance.

### 3.6 Expected performances and trade-off studies for the choice of the baseline orbit

To assess the performances of the payload to detect GRB two elements are need: 1) the response of the instrument, including on axis and off axis effective areas and redistribution matrices [26]; 2) a GRB mock population. We used the mock population produced and developed by [33, 35]. The free parameters of the population are constrained by comparison with well-known and well studies GRB catalogues, such as the Swift/BAT and Fermi/GBM catalogues. Mock GRBs are distributed isotropically in the sky, and for each event in the mock population, the spectrum, assumed to be represented by a Band function, is convolved with the effective area corresponding to its angular position. Figure 3 shows the locus of the HERMES pathfinder detected GRB in the $E_{peak}$-flux plane. The HERMES population is compared with the one actually detected by Fermi/GBM. We note that the majority of the Fermi/GBM GRBs should be detected also by HERMES pathfinder. We miss some of the faintest/hardest GRBs, but gain some soft GRBs (thanks to the broad band extending to soft energies, inaccessible to the Fermi/GBM). More detail will be presented in [36].

Each individual HERMES pathfinder payload should be able to detect >150 GRBs per year, ~10% of which should be short GRBs [36]. Of course, the number of GRB observed by at least three satellites is lower than this figure, and strongly depend on the optimization procedures [27]. Once a GRB is observed by three or more satellites its position in the sky can be reconstructed using the arrival delay time of the photons on different detectors. [14] presents a detailed analysis of the localization performances expected using the optimizations at the basis of the [27] simulated mission analysis. This is aimed at maximizing the total number of GRBs observed simultaneously by at least three satellites with a minimum projected baseline higher that a reference value. Increasing this value would provide less GRB observed simultaneously but would also provide longer projected baselines and therefore better position accuracies. Different optimization strategies will be adopted during the mission. At the beginning we must prove our ability to routinely detect GRBs, later we will focus on obtaining the best positions, and therefore on the study of the residual systematic errors adopting the longest possible projected baselines.

Scientific performances also depend on the baseline orbit parameters and on the in-orbit injection strategies. The figures presented above assume a 500 km or 550 km altitude, nearly equatorial orbit, which minimizes the particle background, but also minimizes the background variations, making easier GRB triggers, and maximizes the efficiency of observations, because of the shallower and shorter passages through the South Atlantic Anomaly. The estimated duty cycle in these orbits is between 70% and 80%. The duty cycle on orbits with higher inclinations is lower than this, reaching a minimum of 30–40% on orbits with inclination between 30° and 80°[37]. The efficiency in Sun Synchronous Orbits (SSO) is of the order of 50% because of the high electron fluxes near the geographical poles. Lower altitude orbits have higher observation efficiencies but shorter lifetime due to the orbital decay.

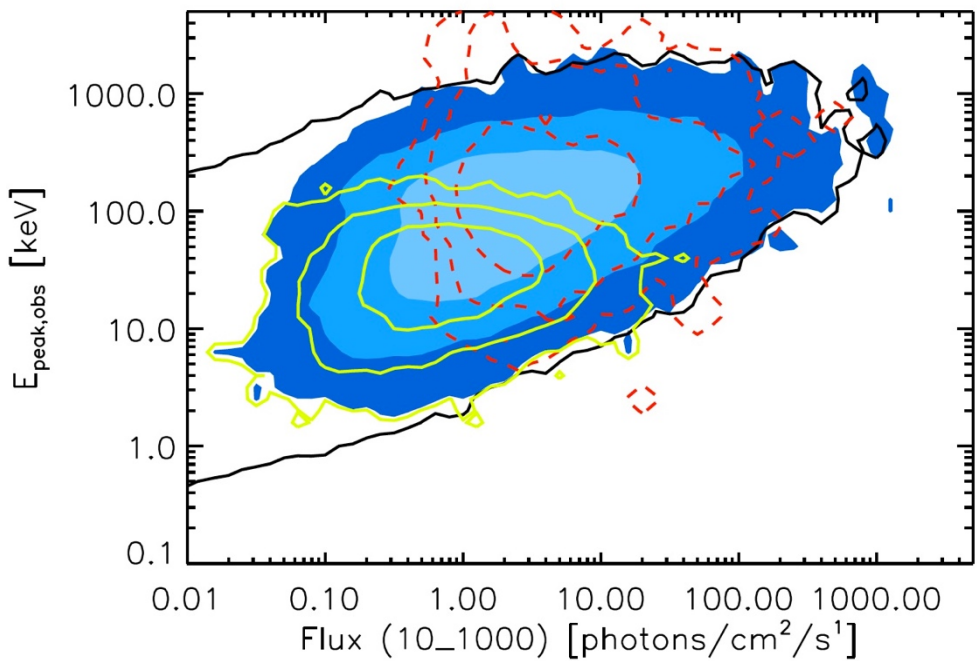

Figure 3. GRBs detectable with HERMES pathfinder in the $E_{peak}$ – observed flux plane (dark to light blue). The solid black line represents the entire simulated population. The green contours highlight the subsample of GRBs detected only with the X instrument. The dashed red lines show the distribution of the real Fermi GRBs in the same plane.

The radiation environment can also directly affect payload performances. Radiation impacting on silicon detectors such as the SDDs used in the HERMES pathfinder payload may change the arrangement of the atoms in the crystal lattice creating lasting structural damage. This in turn creates defect energy levels in semiconductors, which can act as trapping and recombination centers. Damage in the bulk of silicon devices crystal structure is known as 'displacement damage' and results in a net increase of leakage current over exposure time. The maximum allowable leakage current to ensue nominal payload activity is ~500pA. Above this threshold, the low energy threshold increases toward high energies and the electronics becomes less and less able to follow the rate of noise events generated by the detector. We estimated the cumulative dose received by the detectors as a function of time and thus the increment in the leakage current level reached by the detectors in different orbits finding that orbits with height between 500km and 600km and inclination <20° satisfy both the requirements in terms of mission lifetime (2 years), high efficiency and low background, see [25] for details.

### 3.7 Ground Segment

The HERMES-TP/SP ground segment is based on three main components: 1) ground stations (GS), 2) mission operation center (MOC); scientific operation center (SOC). All components shall provide a high level of automation of the nominal mission operations, to reduce the operational cost. This is important for the Pathfinder mission, but it is even more critical for the HFC. Manual operation of such large constellation would be very man-power demanding and impractical.

Downloading 0.5−1 Gbit/day per satellite is not an easy task. The efficiency of CubeSat S-band transmission is of the order of 0.5−1 Mbit/s, thus, to download 1 Gbit of data, about 16−32 m of contact per satellite per day are needed. This is about 3−4 passages per day per satellite, assuming an average time per passage of about 6-8 m, that is 18−24 passages per day for the full mini-constellation of six satellites. This is quite demanding, and requires 1-2 dedicated ground stations. To this purpose, ASI is equipping its Malindi (Kenya) facility with a HERMES pathfinder dedicated ground station. Discussions are on-going with the University of Tasmania to install a second ground station dedicated to HEA CubeSats in the site of Katherine, North Territories, (Australia) where a 12m antenna dedicated to radio astronomy observations and services is already operative.

The MOC and the SOC are realized in the framework of the HERMES-SP project. The following main functional components are typically identified within the MOC: 1) Mission Control System (MCS); Mission Planning System (MPS), Flight Dynamics System (FDS), Mission Automation System (MAS). The MCS will provide the mechanisms to monitor the health of the space segment based on real-time and/or back orbit telemetry data received on the ground. It will also provide the means to command the space segment. The MCS will be in charge of receiving and distributing the scientific telemetry data to the SOC. The MPS will provide the capabilities to plan all data collections and any relevant activities required to control the space segment operation for the constellation. It is expected that then operations will be based on a reference plan that will be built based on the MPS mission configuration and on inputs provided by the SOC. The FDS will provide the mechanisms to perform any required space segment orbit monitoring activities based on GPS information from the on-board equipment. Since the satellite does not provide any mechanisms to perform orbit correction manoeuvres, the capabilities of the FDS will be limited to monitoring functions. The MAS will be in charge of the management of the automation infrastructure used for the execution of the automatic flight operations.

The data produced by each payload and the housekeeping data produced by each SM will be transmitted by the MOC to the SOC, which will be hosted by ASI Space Science Data Centre. The tasks to be carried out by the SOC are the following: 1) Monitoring and health checks of the payload and of the scientific data; 2) Low level scientific data reduction, including conversion of telemetry in FITS format, run pipelines of scientific data reduction (linearization, cleaning etc.), verification of the triggers sent by the different payloads in response to a GRB; 3) high level data analysis, determination of the position of the GRB and of the relative uncertainties through the measured delays in the arrival time of the signals on different detectors; 4) validation of GRB spectra, light curves, delays and positions; 5) archive and dissemination of raw and reduced data, including positions. Particular care is placed on the study of the statistical and systematic effects that limit the estimation of the time delays in the arrival of the signals on different instruments, and therefore in the determination of GRB positions in the sky [see 14 for details].

### 3.8 Programmatic status

Both payload and service module critical design review were successfully passed on October/November 2020. Production and test (both functional and environmental) of the first Proto Flight Module (PFM) is foreseen for Q1/Q2 2021. Production and test of the other five Flight Modules (FMs) will follow, ending Q4 2021. SOC & MOC development and test will also be completed by Q4 2021. SOC will be hosted by ASI SSDC. Orbit injection of the six-satellite mini-constellation should occur in the second half of 2022. The first few months of mission will be dedicated to commissioning of both SM and payload and to payload in flight calibrations, which will be carried out with a few days' long observations of the Crab pulsar.

## 4. ENVIRNOMENT & ECOLOGY

HERMES-TP/SP and SpIRIT are not unique in the international panorama. On the contrary, it is part of a fertile environment that is today extremely rich in projects, ideas, proposals. Since GRBs are relatively bright, more teams around the world are planning to monitor them using nano-satellites and constellations of nano-satellites. We list in the following the main on-going projects.

*BurstCube*, a 6U CubeSat developed by NASA which will detect GRBs using four CsI scintillators, each with an effective area ~90 cm$^2$ [38]. BurstCube is expected to be launched at the end of 2021.

The *Educational Irish Research Satellite 1* (EIRSAT-1), supported by ESA's *Fly your satellite program*, will carry a gamma-ray module (GMOD) to detect gamma-ray bursts [39]. GMOD uses SensL B-series SiPM detectors and a CeBr scintillator. EIRSAT-1 will be launched from the ISS in 2021.

Nanosatellite constellations include the Chinese *Gamma-Ray Integrated Detectors* (GRID), which will consist of GRB detectors (as secondary payloads) on 10–24 CubeSats [40].

The *Cubesats Applied for MEasuring and Localising Transients* (CAMELOT) constellation is expected to consist of at least 9 nano-satellites [41]. The in-orbit demonstration of the GRB detector for the CAMELOT constellation will fly as a secondary payload on the 3U CubeSat VZLUSAT-2 in the beginning of 2021. Another in-orbit demonstration mission for CAMELOT will fly on a 1U CubeSat GRBAlpha in the first half of 2021 [42].

Coordination activities of these (and other proposed) projects are actively on going in the framework of the GRB nano-satellite list[1].

The high scientific relevance of high-energy transient localization and study, and the crucial synergy with advanced LIGO/VIRGO observations in the next few years, prompted the preparation or the finalization of three observatories dedicated to transient astrophysics in China: GECAM [43], SVOM [44] and Einstein Probe [45]. GECAM has been successfully launched on a 600 km LEO on December 10 2020. It consists of two micro-satellites in LEO hosting each one an array of 25 LaBr$_3$ scintillator crystals read out by SiPMs, sensitive in the broad 5–1500 keV band. The orbital configuration should ensure instantaneous all-sky coverage. GECAM GRB localization accuracy should be the order of 1 or a few degrees. SVOM is a French/Chinese collaboration for a GRB mission to be launched in 2021. It hosts both wide field and narrow field instruments, the ECLAIRs instrument is a coded mask telescope similar to SWIFT/BAT in concept, with similar FoV, similar energy band 10–250keV, and about 1/4 collecting area. It can provide GRB positions with accuracy down to 15 arcmin. Einstein Probe is also scheduled for launch in 2021. It hosts a micro-pore, lobster-eye telescope with FoV of about 1 sr and energy bandpass 0.5–4 keV, thus nicely complementing the hard X-ray instruments on board GECAM and SVOM. It can provide source localization with arcmin accuracy. The three Chinese observatories should be active at the time HERMES-TP/SP (and the other CubeSat missions listed above) will fly, as well as the currently active observatories for transient science (Swift, Integral, Fermi and Agile). It would therefore interesting and useful to study synergies and implement collaborations among this impressive fleet of space-based observatories, to both enhance the scientific exploitation of transient science and best prepare for the next generation sensitive high-energy all sky monitor.

**Acknowledgments**

This work has been carried out in the framework of the HERMES-TP and HERMES-SP collaborations. We acknowledge support from the European Union Horizon 2020 Research and Innovation Framework Programme under grant agreement HERMES-Scientific Pathfinder n. 821896 and from ASI-INAF Accordo Attuativo HERMES Technologic Pathfinder n. 2018-10-H.1-2020.

---

[1] https://lists.nasa.gov/mailman/listinfo/grbnanosats